\RequirePackage{fix-cm}
\documentclass[twocolumn,natbib]{svjour3}          

\smartqed  
\usepackage{graphicx}
\usepackage{color}
\usepackage{amsmath}
\usepackage{amssymb}

\usepackage{mathptmx}      
\newcommand{\bphi}{\boldsymbol{\phi}}

\begin{document}

\title{Inference of epidemiological parameters from household stratified data}


\author{James N.~Walker \and
        Joshua V.~ Ross \and
        Andrew J.~Black}


\institute{
School of Mathematical Sciences,\\
University of Adelaide,\\
Adelaide,\\
SA 5005,\\
Australia\\
              \email{andrew.black@adelaide.edu.au}   
}

\date{Received: date / Accepted: date}

\maketitle

\begin{abstract}
We consider a continuous-time Markov chain model of SIR disease dynamics with two levels of mixing. For this so-called stochastic households model, we provide two methods for inferring the model parameters---governing within-household transmission, recovery, and between-household transmission---from data of the day upon which each individual became infectious and the household in which each infection occurred, as would be available from first few hundred studies.
Each method is a form of Bayesian Markov Chain Monte Carlo that allows us to calculate a joint posterior distribution for all parameters and hence the household reproduction number and the early growth rate of the epidemic. The first method performs exact Bayesian inference using a standard data-augmentation approach; the second performs approximate Bayesian inference based on a likelihood approximation derived from  branching processes. These methods are compared for computational efficiency and posteriors from each are compared. The branching process is shown to be an excellent approximation and remains computationally efficient as the amount of data is increased.

\end{abstract}


\section{Introduction}



First few hundred (FF100) studies are data collection exercises carried out in the early stages of pandemic influenza outbreaks \citep{ADHA:2014,HPA:2009,McLean:2010,Gag-Laf:2012}. The aim of these is to characterise a novel strain to determine its impact and hence inform public health planning \citep{McCaw:2013,Read:2013}. FF100 studies involve the collection of data from households where one person is confirmed to be infected. The members of the household are surveilled to identify their time(s) of symptom onset and the study is continued until the first few hundred cases have been observed, or adequate characterisation has been achieved. Households are the primary unit of observation because they are convenient to surveil---in contrast to more general contact tracing---and a large fraction of transmission occurs within the household \citep{Ghani:2009}.  \\

Stochastic models, where the population are split into households with different rates of mixing within and between households, are a natural framework to understand such data \citep{Ball:1997}. Recent work inferred \emph{within-household} epidemic parameters from this type of household stratified data \citep{Ball:2015,Black:2016}. In \citet{Black:2016}, inference is performed using a Bayesian MCMC framework, with exact evaluation of the likelihood, returning a joint posterior distribution for all parameters of interest and hence the within-household reproductive ratio.
In this paper we present a new method for performing inference for a Markovian SIR household model---that also infers the between household transmission parameter---and compare it to a standard approach. With an estimate for the between household mixing we can then in turn estimate the household reproductive number, $R_*$, and the early growth rate of the epidemic, $r$, which are of importance to public health response \citep{McCaw:2013,Read:2013}.\\


\begin{figure}[ht!]
	\centering
	\includegraphics[width=\columnwidth]{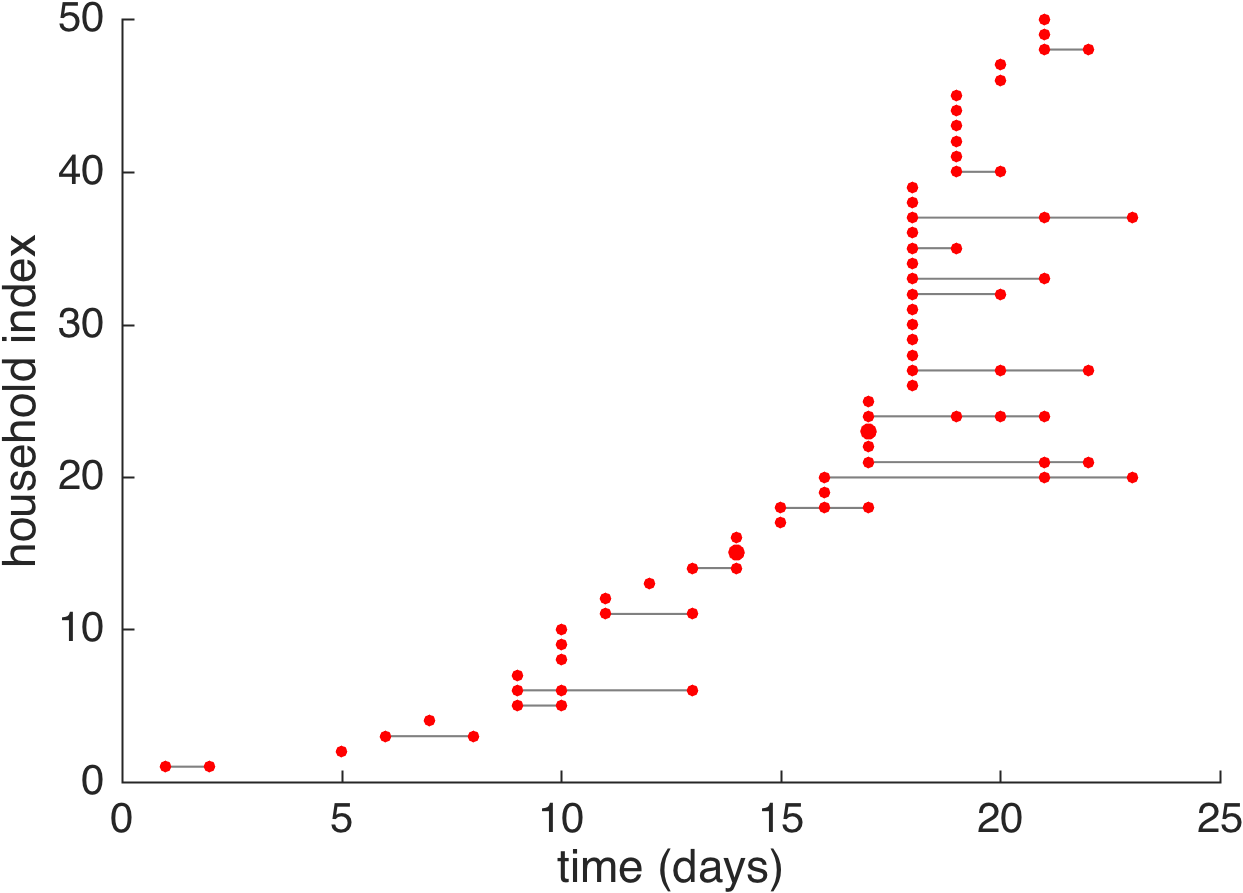}
\caption{A realisation of the SIR household model (described in Section \ref{sec:model}) with households of size 3. The times of symptom onset, binned into days, in the first 50 infected households at the beginning of an epidemic outbreak are presented. The size of points corresponds to the number of infections on that day. The lines provide a visual reference to link infections within the same household.}
\label{fig:ts}
\end{figure}

The data we assume to be available are illustrated in Figure \ref{fig:ts}; we observe only the times, at a daily resolution, when individuals become symptomatic, which is assumed to coincide with infectiousness; recovery times are not available. 
The main challenge in this, as with many similar models, is that the likelihood is difficult to compute due to the missing data. The standard approach to these sorts of problems is to use a data-augmentation method \citep{ONeill:1999,Demiris:2005}. In this approach, all unobserved events are treated as unknowns to also be sampled within the MCMC routine; for the model considered in this paper these would be the exact infection and recovery times for each individual within each household. When the exact times are known the likelihood is trivial to evaluate. A data-augmented approach potentially allows great flexibility in model choice and fitting, but the trade off is that the MCMC scheme needed to sample from the joint distribution of parameters and unknown data is quite complex and needs to run for longer to achieve proper mixing. Convergence can be an issue when there is a large amount of missing data \citep{McKinley:2014,Pooley_2015} and the scalability of these algorithms is poor as more data is added \citep{Cauchemez:2008}---DA MCMC is essentially a serial algorithm that works on the whole data set at once and cannot exploit parallelism easily. \\

Here we develop another approach, based on approximation of the original process, and compare it to a data-augmentation method. Our approach is to carefully consider the dynamics and structure of the problem to allow us to derive an approximation to the exact likelihood that can be evaluated using a novel combination of numerical methods (matrix exponential methods \citep{Sid98}, stochastic simulations \citep{Gillespie:1976} and numerical convolutions). This allows us to use a simple Metropolis-Hastings algorithm to compute a joint posterior for all the parameters of interest. There are two main assumptions underpinning our method. The first is that we can approximate the early time behaviour of the epidemic as a branching process where only a single introduction to each household is possible. This is a very mild assumption and we would expect data collected in the early stages of an outbreak, say from an FF100 study, to conform to this reasonably closely. The second, more technical assumption we make, is that we can replace certain random variables that arise in the problem with their mean values. 
We show that our method provides a very good approximation to the full model and the final posteriors that we compute show good convergence to the true model parameters as the amount of household data is increased.

\subsection{Overview of the paper}

The rest of the paper is organised as follows. The Markovian household model and the data we assume is available is introduced in Section \ref{sec:fullmodel}, and the data-augmented MCMC method is briefly discussed in Section \ref{sec:DAMCMC}. The branching process approximation and associated threshold quantities are introduced in Section \ref{sec:model}. We show how the likelihood can be decomposed into two parts related to the within- and between-household dynamics. 
In Section \ref{sec:between} we use an approximation to allow us to calculate the likelihood of seeing a given number of newly infected households over a day. This calculation depends on evaluating a number of conditional expectations relating to the dynamics of a single household.
These calculations are detailed in Section \ref{sec:within} along with calculation of the part of the likelihood due to the within-household dynamics. Section \ref{sec:results} shows our results in the form of posterior distributions for the parameters of the model for both methods and demonstrates how these converge 
as more data is included from longer observations of the epidemic. The methods are compared in terms of the similarity of results as well as their efficiency. We conclude the paper in Section \ref{sec:discussion} with a discussion of our methods, their weaknesses and possible extensions.


\section{Households model and data}
\label{sec:fullmodel}
The dynamics of the epidemic are modelled as a continuous-time Markov chain.
Individuals are grouped into  $H$  mutually exclusive households and make effective contact at a high rate within households and at a low rate between households.
In this paper, for simplicity, we will assume that all households are of the same size, $N$, and an SIR model for disease dynamics. Thus each individual is classified as \emph{susceptible} to infection, $S$, \emph{infectious} and able to infect susceptible individuals, $I$, or \emph{recovered} and immune to the disease, $R$. As $N$ is fixed, the state or configuration of a household can be specified by the number of susceptible and infectious individuals within the household (where $R=N-S-I$). 

If we index households by $j=1,\dots,H$, then the sate of the system, $Y(t)$ can be specified by a $H\times2$ matrix where the $j$'th row gives the number of $S$ and $I$ in household $j$,
\begin{equation}
	Y(t) = \left(s_j(t),i_j(t) \right)_{j=1:H}.
\end{equation}
Thus the state space is then (dropping the dependence on time),
\begin{equation*}
\mathcal{S}=\left\{(s_j,i_j)_{(j=1:H)} \in \{0,1,\dots,N\}^{H\times 2}  \, | \,  s_j+i_j\leq N \;\; \forall \;\; j  \right\}.
\end{equation*}
Note that there are lower dimensional representations of household models in which a state is a vector which describes the total number of households in each possible configuration \citep{Black:2014}; however, we adopt the higher dimensional version here as it simplifies inference.\\

The dynamics of the SIR household model are defined by the transitions that can occur throughout the population and their corresponding rates. Infectious individuals make effective contact within their household at rate $\beta$. In household $j$ the probability that effective contact within the household leads to an infection is $\frac{s_j}{N-1}$, and hence the rate of within household infection is $\frac{\beta s_j i_j}{N-1}$. Each infectious individual recovers at rate $\gamma$, so recoveries in household $j$ occur at rate $\gamma i_j$. Lastly, infectious individuals may make effective contact with any individual in the population outside of their own household at rate $\alpha$. Thus between household effective contact results in an infection in household $j$ at rate 
\begin{equation*}
  \frac{\alpha s_j(I-i_j)}{N(H-1)},   
\end{equation*}
where $I=\sum_j i_j$ is the total number of infectious individuals in the population.\\

We assume that the first infection is seeded in a single household at some U$(0,1)$ distributed time, $\theta_0$, such that the matrix encoding the first state, $Y(\theta_0)$, has first row $(N-1,1)$ and all other rows $(N,0)$. Note that, as our data only reveals cases of infectiousness at a daily resolution, the time of the first infection is unknown.\\

\subsection{Data}
\label{sec:data}

Suppose we have observed the start of an epidemic over some time period $(0,T]$. We assume our data counts the cumulative number of infections in each household each day, where day $t$ is defined as the time interval $(t-1,t]$. Here we are assuming that symptoms coincide with infectiousness. 
Each household is labelled by $j=1,\dots, M$ in the order that they became infected, but note that as the process is only observed at a daily resolution (taken to be the end of each day) the ordering within a day is arbitrary.
It is natural to specify this data in terms
of two quantities: the days on which each household is infected and
the time series of cumulative infection counts within each household,
starting from their day of infection. More precisely, let $\psi_t$ be
the set of the labels ($j$) of the households that became infected on
day $t$. Then let $\mathbf{w}^{(j)}=(w_k)^{(j)}$ be a vector where
$w_k$ is the cumulative number of infection events within the $j$th
household, recorded at the end of day $k$, from the day of the households initial infection up to day $T$.  Thus the
data is completely specified by the sets $\{\psi_t\}_{t=1:T}$ and the
vectors $\{\mathbf{w}^{(j)}\}_{j=1:M}$, which we denote
\begin{equation}
     \mathcal{D} = \left\{ \{\psi_t\}_{t=1:T} , \{\mathbf{w}^{(j)}\}_{j=1:M} \right\}.
\end{equation}

\begin{figure}[tb]
	\centering
	\includegraphics[width=\columnwidth]{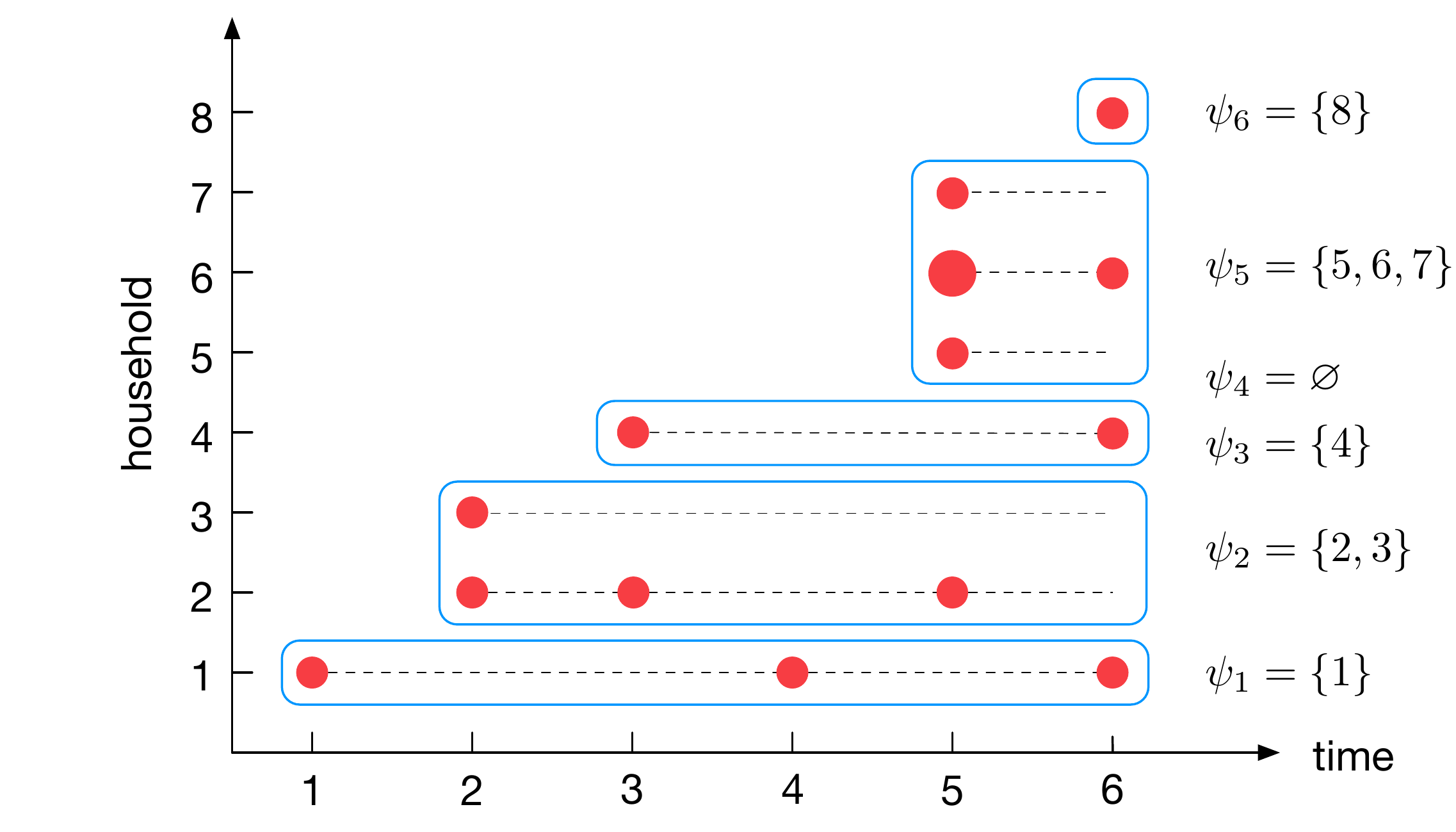}
	\caption{An illustration of how the data is structured for
          inference. An outbreak observed over $T=6$ days, resulting
          in $M=8$ households becoming infected. The red circles
          indicate the days on which new infections are observed and
          their size is proportional to the number of infections. The sets $\psi_t$
          indicate which households become infected on day $t$. Note
          that $\psi_4 = \varnothing$ indicates that no new houses
          were infected on day 4. The cumulative number of observed
          cases within each household, over the 6 days are:
          $\mathbf{w}^{(1)}=(1,1,1,2,2,3)$,
          $\mathbf{w}^{(2)}=(1,2,2,3,3)$,
          $\mathbf{w}^{(3)}=(1,1,1,1,1)$,
          $\mathbf{w}^{(4)}=(1,1,1,2)$, $\mathbf{w}^{(5)}=(1,1)$,
          $\mathbf{w}^{(6)}=(2,3)$, $\mathbf{w}^{(7)}=(1,1)$ and
          $\mathbf{w}^{(8)}=(1)$.}
	\label{fig:data}
\end{figure}

 These quantities are
illustrated for a specific example in Figure \ref{fig:data}.  We also
define $\Omega_t = \cup_{j=1}^{t-1} \psi_j$, which is the set of
labels of households that became infected before day $t$; this will be used in the
branching process approximation in Section \ref{sec:model}.\\

\section{Data augmented MCMC}
\label{sec:DAMCMC}
Data augmented Markov Chain Monte Carlo (DA MCMC) is a powerful, exact Bayesian inference method for data with missing information. We adopt an approach similar to \cite{ONeill:1999} to infer the joint posterior density of $(\alpha,\beta,\gamma)$. The general approach is to construct an augmented likelihood, the joint density of the data and the missing information given the model parameters, and use this to construct a single-component Metropolis-Hastings algorithm. This method proves useful for FF100 study data as the exact times of infection over each day are missing and the number of recovery events, and the times at which they occur, are entirely unknown. Although the data-augmented approach is a standard method for this kind of problem, we are not aware of it having been implemented in a household model where neither recovery or infection times are observed exactly and in which all parameters are unknown. 
For example, data-augmented MCMC has been implemented for a similar model with data obtained at regular discrete times, however parameters associated with the infectious period distribution were assumed to be known \citep{Demiris:2014}. Hence we outline the algorithm used for our particular problem.\\


 As per the usual approach we augment our data with the transition times $\boldsymbol{\theta}\in \mathbb{R}^m$ and corresponding states $\mathbf{Y} = \{Y(\theta_1),\dots,Y(\theta_m)\}$ in the underlying model, where $m$ is the unknown number of transitions over time $(\theta_0,T]$ which is allowed to vary. Additionally we consider the classification of infection events as missing, that is, we augment the data by transition labels $\zeta\in\{\text{recovered, within, between}\}^m$. This is such that we can construct sets of transition indices, $A$, $B$ and $C$, which correspond to within-household infection, between-household infection and recovery events respectively. In writing down the expression for the augmented likelihood function we adopt the convention that all quantities ($s,i$) are evaluated immediately prior to a transition. Hence we have,
 \begin{align*}
L_{DA}:&=f\left(\mathcal{D},\theta,\mathbf{Y},\zeta|\alpha, \beta,\gamma,\theta_0 \right)\\
 &=1_{\{\mathcal{D},\theta,\mathbf{Y},\zeta \}}
 \prod_{j\in A}\frac{\beta s^{(j)}i^{(j)}}{N-1}
 \prod_{k\in B}
 \frac{\alpha s^{(k)}\left(I-i^{(k)}\right)}{N(M-1)} 
 \prod_{l\in C}\gamma i^{(l)}\\
& \hspace{-0.5cm} \times \text{exp}
\left
\{-\sum_{p=1}^{m+1}\sum_{c=1}^{H}
\left(\frac{\beta s_{c}^{(p)}i_{c}^{(p)}}{N-1}
+\frac{\alpha s_{c}^{(p)}\left(I-i_{c}^{(p)}\right)}{N(M-1)}
+\gamma i_c^{(p)} \right) \right.\\
 & \hspace{5cm} \times(\theta_p-\theta_{p-1}) \Bigg \},
\end{align*}
where $1_{\{\mathcal{D},\theta,\mathbf{Y},\zeta \}}$ denotes an indicator function corresponding to one if the data, $\mathcal{D}$, could have arisen from the events defined by ($\boldsymbol{\theta}$, $\mathbf{Y}$, $\zeta$), and $\theta_{m+1}:=T$ for simplicity.
Note that inference could be made without labelling the two kinds of infection, however this more explicit representation produces gamma or truncated gamma marginal densities of $\beta$ and $\alpha$ for uniform, gamma, inverse uniform or truncated gamma priors; hence they may be efficiently sampled. \\

Marginal posterior densities of $\alpha$, $\beta$, $\gamma$ and $\theta_{0}$ can be evaluated and sampled from in a similar way to \citep{ONeill:1999}. Lastly the density, $f\left(\boldsymbol{\theta},\mathbf{Y},\zeta | \beta,\gamma,\theta_0,{\mathcal{D}}\right)$, is proportional to $L_{DA}$, thus it can be sampled from by randomly choosing from the following five kinds of moves according to an arbitrary probability mass function with non-zero components, $\{q_1,\dots,q_5\}$:\\ 

\noindent (i) Randomly select an infection time, $\theta^{(j)}$, choose a candidate $\text{U}\left(\lfloor \theta^{(j)}\rfloor,\lceil\theta^{(j)}\rceil\right)$ distributed infection time. Let the augmented likelihood corresponding to the candidate be denoted by $\hat{L}_{DA}$. The new point is accepted with probability
\begin{equation}
\text{min}\left\{\frac{\hat{L}_{DA}}{L_{DA}},1\right\};
\label{eq:exc_prob}
\end{equation}
(ii) Randomly select an infection event and change its type, $\zeta^{(j)}$, from between to within household infection or vice versa. The new point is accepted with probability as in Eq.~\eqref{eq:exc_prob};\\

\noindent
(iii) Randomly select a recovery time, $\theta^{(j)}$, and choose a candidate $\text{Uniform}\left(\theta^{(k)},T\right)$ distributed recovery time, where $\theta^{(k)}$ is the time of the first infection within the household. The new point is accepted with probability as in Eq.~\eqref{eq:exc_prob};\\

\noindent
(iv) Insert a $\text{Uniform}\left(\theta^{(k)},T\right)$ distributed recovery time in a randomly chosen household. Let $M$ be the number of households infected by time $T$. The new point is accepted with probability
\begin{equation*}
\text{min}\left\{\frac{\hat{L}_{DA}M\left(T-\theta^{(k)}\right)q_5}{L_{DA}(|C|+1)q_4},1\right\};\text{ or},
\end{equation*}
(v) Randomly select and remove a recovery event with probability
\begin{equation*}
\text{min}\left\{\frac{\hat{L}_{DA}|C|q_4}{L_{DA}M\left(T-\theta^{(k)}\right)q_5},1\right\}.\\
\end{equation*}

Each iteration of the DA MCMC algorithm is comprised of Gibbs samples of $\alpha$, $\beta$, $\gamma$ and $\theta_0$ followed by a Hastings step for $(\boldsymbol{\theta},\mathbf{Y},\zeta)$ as per (i)-(v). The distribution of these samples converge to the posterior distribution of $(\alpha,\beta,\gamma)$, though consecutive samples will be highly correlated.


\section{Branching process approximation}
\label{sec:model}
An alternative approach to inference is to assume that our model acts like a branching process at the household level. That is, we assume an infinite population size and hence that between-household infection only occurs into completely susceptible households; this assumption is reasonable, as the data we wish to perform inference on is from the very earliest stages of an outbreak. Households act independently after initial infection, hence we can consider the dynamics within each infected household, following their initial infection, in isolation from each other \citep{Ball:1997,BHKR12}. Under this model we construct an approximate likelihood with the aim of obtaining accurate estimates for the joint posterior distribution of $(\alpha,\beta,\gamma)$. We show in Section \ref{sec:results} that the resulting distribution approximates the exact distribution very well, while the independence assumption allows for computational gains in the inference as the data set grows in size.\\

As we are considering households in isolation from each other, we define the state space for a single household as 
\begin{equation}
     \mathcal{S}=\{(s,i)\in\{0,1,...,N\}^2: s+i \le N\}. \nonumber
\end{equation}
The within-household dynamics are defined by the transitions that can occur within an individual household and their corresponding rates; these are simply the within-household infection and  recovery transitions with rates as described in Section \ref{sec:fullmodel}. The within-household process can be defined in terms of its infinitesimal transition rate matrix, $Q$, given by
\begin{equation*}
[Q]_{f(s,i),f(x,y)} = \left\{
     \begin{array}{lr}
      \frac{\beta s i}{N-1} & \text{for }(x,y)=(s-1,i+1)\text{, } s\geq1 \\
       \gamma i & \text{for }(x,y)=(s,i-1)\text{, } i\geq1\\
       -\frac{\beta s i}{N-1}-\gamma i &\text{for }(x,y)=(s,i) \\
       0 & \text{otherwise,}
     \end{array}
   \right.
\end{equation*}
where $f: \mathcal{S}\rightarrow \{1,...,|S|\}$ is a bijective map. The first infection within each household moves it into state $(N-1,1)$, at which point the within-household dynamics determine how the disease spreads within the household for the remainder of the epidemic. \\

We assume that between-household infection occurs due to homogeneous mixing of all the individuals in the population at rate $\alpha$, thus the rate at which new households are infected is simply $\alpha I(t)$, where $I(t)$ is the total number of infected individuals in the population at time $t$. The model is initialised with a single infected household at a $\text{U}(0,1)$ distributed time. \\

For this model we can identify the threshold parameter, $R_*$, which is a household (population level) reproduction number \citep{Ball:1997}. 
This is the expected number of households infected by a primary infectious household in an otherwise susceptible population of households; where a household is considered infectious while it contains at least one infectious individual and a household is considered susceptible if it contains only susceptible individuals. It is one of at least five reproductive numbers that might be used when assessing the controllability of a disease in a community of households \citep{Pellis09,Pellis15,Goldstein09}, but we adopt it herein as it is relatively easy to calculate and interpret. Let $\{X_t\}_{t\in\mathbb{R}^+}$ be the Markovian process that describes the state of an individual household from the time of its infection (i.e., the time of the first infection within the household). Let $I(k)$ be the function which returns the number of infectious individuals corresponding to state $k$. Then we have
\begin{equation}
     R_*=E\left[\int_0^{\infty}\alpha I(X_t)\ dt \right],   \nonumber
\end{equation}
where $X_0 = (N-1,1)$ is the initial state of the process \citep{Ball:1997,Ball99,RHK10}. This can be calculated by solving a system of linear equations that depend on the parameters of the epidemic model \citep{RHK10,PS02}. \\

Also of interest from a public health perspective, is the early growth rate, $r$; this is also called the Malthusian parameter. Under the same conditions as above, this is defined as the unique solution to
\begin{equation}
E\left[\int_0^{\infty}\alpha I(X_t) e^{-rt}\ dt \right] = 1.\nonumber
\end{equation}
This can once again be evaluated efficiently~\citep{RHK10}.

\subsection{Approximate likelihood}
\label{sec:datalike}

The branching process likelihood approximation relies on expressing the likelihood in terms of the data on a given day, $t$, partitioned into newly infected households, $\psi_t$, and formerly infected households each day, $\Omega_t = \cup_{j=1}^{t-1} \psi_j$ (see Section \ref{sec:data}). With this partition, the likelihood for $(\alpha,\beta,\gamma)$ can then be written as,
\begin{equation}
	L(\alpha,\beta,\gamma) =\prod_{t=1}^T P\left(\{\mathbf{w}^{(j)}\}_{j\in
          \psi_t} \Big\vert \psi_t \right) P\left(\psi_t \Big\vert \{
        \mathbf{w}^{(j)}\}_{j\in \Omega_t} \right);
\label{eq:like1}
\end{equation}
we have invoked the independence between
$\left(\{\mathbf{w}^{(j)}\}_{j\in \psi_t} \vert \psi_t\right)$ and
$\{\mathbf{w}^{(j)}\}_{j\in \Omega_t}$ due to the branching process
assumption. Note that there is also a more subtle assumption of independence in this factorisation of the likelihood: even in the branching process approximation, the infection of new households contains information about the within-household processes -- that there is at least 1 individual still infectious within the $\Omega_t$ households -- but there is no easy way of incorporating this into the likelihood; in any case, the contribution to the likelihood of this will be small after more than two households have become infected.\\

As $\Omega_1 = \varnothing$, that is, there are no
households infected before $t=0$, the term
\begin{equation}
P\left(\psi_1 \Big\vert \{ \mathbf{w}^{(j)}\}_{j\in \Omega_1} \right)
= P\left(\psi_1\right)\nonumber
\end{equation}
is determined by the initial condition. Further, households in
$\psi_t$ are identically distributed in the absence of within-household information. Thus their labels are arbitrary and only the
number of households in $\psi_t$ is relevant, that is
\begin{equation*}
P\left(\psi_t \Big\vert \{ \mathbf{w}^{(j)}\}_{j\in \Omega_t} \right)
= P\left(|\psi_t| \Big\vert \{ \mathbf{w}^{(j)}\}_{j\in \Omega_t}
\right).
\end{equation*}
Thus we can factor the likelihood, Eq.~\eqref{eq:like1}, into two parts that are related to the within-household dynamics and between-household dynamics respectively. That is,
$$L(\alpha,\beta,\gamma)=L_w(\alpha,\beta,\gamma)L_b(\alpha,\beta,\gamma),$$
where 
\begin{equation}
	L_w(\alpha,\beta,\gamma)=\prod_{t=1}^T P\left(\{\mathbf{w}^{(j)}\}_{j\in \psi_t} \Big\vert  \psi_t \right)
	\label{eq:withinhhlh}
\end{equation}
and
\begin{equation}
	L_b(\alpha,\beta,\gamma)=\prod_{t=1}^T P\left(|\psi_t| \Big\vert  \{ \mathbf{w}^{(j)}\}_{j\in \Omega_t}  \right).
	\label{eq:between_like}
\end{equation}
We refer to $L_w$ as the within-household likelihood function and $L_b$ as the between-household likelihood function. In Section \ref{sec:between} we detail how we calculate $L_b$ and in Section \ref{sec:within}, $L_w$.

\subsection{Between-household likelihood, $L_b$}
\label{sec:between}

Each term in the product for the between-household likelihood, Eq.~\eqref{eq:between_like}, is the probability that we observe $H_t:=|\psi_t|$ new infected households on day $t$, given the data, over the time period $[0,T]$, for households that were infected before day $t$.  We decompose $H_t$ into two components, $H_t^{(1)}$ and $H_t^{(c)}$, such that $H_t = H_t^{(1)} + H_t^{(c)}$. The first component, $H_t^{(1)}$, is the number of the newly infected households on day $t$ that are infected by a household in  $\Omega_t$, i.e., a household infected before day $t$. The second component, $H_t^{(c)}$, is the remaining number of newly infected households on day $t$, i.e., those that are infected by households that become infected on day $t$. We do not observe this demarcation in our data, but it assists us in the evaluation of the likelihood.\\

We start by considering the calculation of the probability mass function (pmf) of $H_t^{(1)}$, denoted $\mathbf{h}_t^{(1)}$, in Section~\ref{sec:first_gen}. Then, in Section~\ref{sec:sgh}, we consider the evaluation of the pmf of $H_t^{(c)}$, $\mathbf{h}_t^{(c)}$. The required pmf of $H_t$, $\mathbf{h}_t$, is subsequently evaluated using efficient methods for calculating convolutions.

\subsubsection{First generation of households, $H_t^{(1)}$}
\label{sec:first_gen}

To calculate $\mathbf{h}_t^{(1)}$, the pmf of the number of first generation infected households, we note that on the first day of the epidemic there is only a single household infected at a $U(0,1)$ distributed time. Hence there is exactly 1 infected household in the first generation of households, so 
\begin{equation*}
P\left(H_1^{(1)}=1\right)=1.
\end{equation*}
For $t\geq2$ we consider the rate at which the households in $\Omega_t$ infect new households. 
As we model the outbreak as a branching process, we assume that only completely susceptible households are infected, hence the instantaneous rate of infection at time $s \in (t-1,t]$ from the households in $\Omega_t$ is 
\begin{equation}
	\alpha \sum_{j\in\Omega_t} I  \left(X^j_s \right), \nonumber
\end{equation}
where $X^j_s$ is the state of household $j$ at time $s$ and $I(k)$ is a function returning the number of infectious individuals in a household in state $k$.\\

Thus the first generation of households are created as an inhomogeneous Poisson process, and conditioning on the information about the households in $\Omega_t$, $\{\mathbf{w}^{(j)}\}_{j\in\Omega_t}$, we have
\begin{equation}
\left(H_t^{(1)} \Big\vert \{\mathbf{w}^{(j)}\}_{j\in\Omega_t}\right) 
     \sim \text{Poisson}\left(\alpha \sum_{j\in\Omega_t}\int_{t-1}^t I\left(X^j_s \Big\vert \mathbf{w}^{(j)}\right) ds \right). 
     \label{eq:full_inhomo}
\end{equation}
Hence, we need to evaluate the distribution of
\begin{equation}
     \Lambda_t:=\alpha \sum_{j\in\Omega_t}\int_{t-1}^tI \left(X^j_s  \Big\vert \mathbf{w}^{(j)}\right) ds.
     \nonumber
\end{equation}
However, this is expensive to compute, so instead we replace $\Lambda_t$ in Eq.~\eqref{eq:full_inhomo} with its expectation, which can be evaluated in a feasible manner. Precisely, we use
$$ 
     P\left(H_t^{(1)}=h \Big\vert \{\mathbf{w}^{(j)}\}_{j\in\Omega_t}\right) \approx \frac{e^{-E[\Lambda_t]}E[\Lambda_t]^{h}}{h!},
$$
where
\begin{equation}
	E[\Lambda_t]= \alpha \sum_{j\in\Omega_t}\int_{t-1}^t E\left[I\left(X^j_s\Big\vert\mathbf{w}^{(j)}\right)\right]ds.
	\label{eq:cond_exp1}
\end{equation}
As $\Lambda_t$ is the force of infection over a short time period (a day) it should have relatively low variance, thus replacing $\Lambda_t$ by its expectation may provide a reasonable approximation. In Section \ref{sec:within} we detail how the conditional expectations, Eq.~\eqref{eq:cond_exp1}, can be calculated using matrix exponential methods. \\

\subsubsection{Subsequent generations of households, $H_t^{(c)}$}
\label{sec:sgh}

Recall that the number of newly infected households on day $t$ is $H_t = H_t^{(1)} + H_t^{(c)}$. The first component, $H_t^{(1)}$, is the number of the newly infected households on day $t$ that are infected by a household in  $\Omega_t$, i.e., a household infected before day $t$. The second component, $H_t^{(c)}$, is the remaining number of newly infected households on day $t$, i.e., those that are infected by households that become infected on day $t$.\\

We assume that the infection of the $H_t^{(1)}$ households are uniformly distributed over day $t$, and since their dynamics are independent, we have that $H_t^{(c)}$ is the convolution of $H_t^{(1)}$ random variables; we will use $G$ to denote one of these random variables. Each of these random variables correspond to the size of a household branching process at time $1$ that was initialised at a U$(0,1)$ time. The calculation of the pmf of $G$ is once again computationally expensive, so here we choose to estimate this distribution using simulation \citep{Gillespie:1976}. As we are simulating over such a short period of time, and use the most efficient representation to minimise computational time, a large number of simulations, producing an accurate estimate, can be produced in a computationally efficient manner.\\


Once we have estimated the pmf of $G$, we can calculate $\mathbf{h}_t$ from $\mathbf{h}_t^{(1)}$ as
\begin{equation}
   \mathbf{h}_t = \mathcal{M} \mathbf{h}^{(1)}_t,
   \nonumber
\end{equation}
where the convolution matrix, $\mathcal{M}$, is defined as follows.
Let $\bphi$ be a column vector of the pmf of the random variable $G+1$. Then let
\begin{equation}
\mathbf{c}_j = \mathbf{c}_{j-1} * \bphi, \quad j\ge2,
\nonumber
\end{equation}
where `$*$' denotes a discrete convolution and $\mathbf{c}_1=\bphi$. The matrix $\mathcal{M}$ is then given by
\begin{equation}
   \mathcal{M} = [\mathbf{e}_1,\mathbf{c}_1,\mathbf{c}_2,\dots],
   \nonumber
\end{equation}
where $\mathbf{e}_1$ is a vector of $0$s with the exception of a $1$ in the first entry. The matrix $M$ is truncated such that no probability needed for the calculation of the likelihood is lost. The calculation of this is not expensive, even for large matrices as the convolutions can be done using using discrete Fourier transforms \citep{Lyons:2011}. In this paper we simply use the built in MATLAB function {\tt conv()}, although other methods may provide computational gains, if required.


\subsection{Single household dynamics, $E[\Lambda_t]$ and $L_w$}
\label{sec:within}

Recall, the evaluation of the pmf $\mathbf{h}_t^{(1)}$ for $t\geq2$, corresponding to the number of first generation infected households on day $t$, requires the evaluation of the expected force of infection over day $t$ from households infected prior to day $t$, $E[\Lambda_t]$. We begin by detailing the evaluation of $E[\Lambda_t]$, and then note how the within-household likelihood, $L_w$, follows.\\


The computation of the $E[\Lambda_t]$ can be expressed in terms of integrals of the expected number of infectious individuals within each household in $\Omega_t$, Eq.~\eqref{eq:cond_exp1}. 
As this expression is a sum over independent households we simplify our exposition, by  detailing the calculation for a single arbitrary household in $\Omega_t$, with observed data $\mathbf{w}$ (thus dropping the superscript `$j$' notation for now). The independence also means we can rescale time within the household to begin at the start of the day of the first infection. Thus we need to calculate the expected number of infected individuals over each of the $|\mathbf{w}|=\omega$ days since the first infection within that house, i.e. 
\begin{align}
E\left[I\left(X_s \Big\vert  \mathbf{w} \right)\right]&=\sum_{k\in \mathcal{S}}I(k)P\left(X_s = k \Big\vert \mathbf{w}\right),
\label{eq:exp_1hh}
\end{align}
for all $s\in(t-1,t]$, where $t=2,\dots,\omega$. Note that as we are conditioning on the entire observed data within the household, $\bf{w}$, the random variable $\omega$ is implicitly conditioned on. That is, we are conditioning on knowing that the household became infected on day $T-\omega+1$. In the remainder of this subsection all probabilities are conditioned on $\omega$, but this is not written explicitly for concision.\\

The expectation Eq.~\eqref{eq:exp_1hh} can be calculated efficiently, and hence the integral of the expectation to find the force of infection can also be calculated efficiently and accurately with Simpsons Rule, say.
Our calculation is similar to that of the forward-backward algorithm \citep{Baum:1970}, but is more involved as we need to calculate the expectations for all $s$, not just the discrete time points at which observations occur.
First we define some quantities. As $w_t$ is the total number of infections observed in the household by the end of day $t$ (within-household time), $X_t$ must be in a set of states such that $N-s=w_t$. These states are encoded by indicator vectors, $\mathbf{z}_t$, with 1s in entries corresponding to states where $N-s = w_t$ and zeros otherwise; these are either row or column vectors as required.\\

Define the row vector $\mathbf{f}_{t}$ as the `forward' probabilities of the system, so the $k$th element is the probability the system is in state $k$ at the end of day $t$, given the observed data up to $t$, 
 \begin{align*}
 [\mathbf{f}_{t}]_k = P\left(X_t=k \vert \mathbf{w}_{(1:t)} \right).
 \end{align*}
These can be calculated in a recursive manner as follows:
\begin{equation}
  \mathbf{f}_t = \frac{(\mathbf{f}_{t-1} e^Q) \circ \mathbf{z}_t}{(\mathbf{f}_{t-1} e^Q) \cdot \mathbf{z}_t}, \quad t=2,\dots,\omega , \nonumber
\end{equation}
where `$\circ$' is an element-wise vector product. The first vector, $\mathbf{f}_1$, is determined from the initial condition; let $\mathbf{v}$ be a probability vector with a 1 in the entry corresponding to state $(N-1,1)$. As infection is introduced into each household at a $U(0,1)$ distributed time on their day of infection, the distribution of $X_1$ is given by
\begin{equation}
 \mathbf{u} = \int_0^1 \mathbf{v} e^{Q(1-s)} ds . \nonumber
\end{equation} 
Conditioning on $w_1$ gives $ \mathbf{f}_1 = \mathbf{u} \circ \mathbf{z}_1 / \mathbf{u} \cdot \mathbf{z}_1$.\\

We also define the `backward' probabilities, $\mathbf{b}_t$, with elements
 \begin{equation*}
 [ \mathbf{b}_t]_{k} = P\left(\mathbf{w}_{(t+1:\omega)} \vert X_t = k \right). 
 \end{equation*}
These are the probabilities of observing the remainder of the data given that the system is in state $k$ at the end of day $t$. These can be calculated in a similar recursive way to the forward probabilities, but working backward from the final observation:
\begin{equation}
  \mathbf{b}_{t-1} = e^Q (\mathbf{b}_{t} \circ \mathbf{z}_t), \quad t=\omega,\dots, 2, \nonumber
\end{equation}
with $\mathbf{b}_\omega = \mathbf{1}$.\\


 
Applying Bayes' theorem to the pmf in Eq.~\eqref{eq:exp_1hh} and using the Markov property we arrive at
\begin{equation}
P\left(X_s=k \vert \mathbf{w} \right)
= 
\frac{P\left(X_s=k\vert\mathbf{w}_{(1:t-1)} \right)P\left( \mathbf{w}_{(t:\omega)} \vert X_s=k\right)} 
{P\left(\mathbf{w}_{(t:\omega)}\vert \mathbf{w}_{(1:t-1)}\right)},
\label{eq:forbackmod}
\end{equation}
for $s \in (t-1,t]$.
Using the law of total probability and the Markov property on the three probability expressions in Eq.~\eqref{eq:forbackmod} gives
 \begin{equation*}
    P\left(X_s=k \vert \mathbf{w}_{(1:t-1)} \right) = \left[\mathbf{f}_{t-1}e^{Q(s-t+1)}\right]_k,
  \end{equation*}
\begin{equation*}
  P\left(\mathbf{w}_{(t:\omega)} \vert X_s = k\right)
   =
\left[e^{Q(t-s)} (\mathbf{b}_t \circ \mathbf{z}_t) \right]_k,
\end{equation*}
and
\begin{equation*}
  P\left(\mathbf{w}_{(t:\omega)}\vert \mathbf{w}_{(1:t-1)}\right)
  =
  \mathbf{f}_{t-1} \cdot \mathbf{b}_{t-1}.
\end{equation*}
Hence Eq.~\eqref{eq:exp_1hh} can be expressed in a vectorised form as 
\begin{equation}
\begin{aligned}
E\left[I\left(X_s \big\vert  \mathbf{w} \right)\right]
=
\mathbf{i}\cdot
\left(
\frac{\mathbf{f}_{t-1}e^{Q(s-t+1)} \circ e^{Q(t-s)} (\mathbf{b}_t \circ \mathbf{z}_t)}
{\mathbf{f}_{t-1} \cdot \mathbf{b}_{t-1}}
\right), \\
\quad t=2,\dots,\omega, \nonumber
\end{aligned}
\end{equation}
where $\mathbf{i}$ is a vector whose elements are the number of infected individuals in each state.\\



This allows us to numerically evaluate $\mathbf{h}_t^{(1)}$; note that all matrix exponential calculations here can be expressed as $\left[e^Q\right]^a$, so we only need to compute the matrix exponential once per parameter set and take powers of the resulting matrix. Further, when numerically integrating, using a symmetric grid about $t-1/2$ allows us to take advantage of the symmetry of $e^{Q(s-t+1)}$ and $e^{Q(t-s)}$, effectively halving the number of times we need to take powers of $e^Q$.\\

\subsubsection{Within-household likelihood, $L_w$}

Using the quantities calculated above, we can also calculate the within-household likelihood, $L_w$, described in Eq.~\eqref{eq:withinhhlh}. Let $\omega^{(j)}$ denote the length of $\mathbf{w}^{(j)}$. Note, under the branching process assumption, infected households act independently of each other, so their within household dynamics following their infection are independent. Hence,
\begin{align*}
L_w(\alpha,\beta,\gamma)&=\prod_{t=1}^T P\left(\left\{\mathbf{w}^{(j)}\right\}_{j\in\psi_t} \Big\vert \psi_t\right)\nonumber\\
&=\prod_{t=1}^T\prod_{j\in\psi_t} P\left(\mathbf{w}^{(j)}\Big\vert j\in\psi_t\right)\nonumber\\
&=\prod_{j=1}^M P\left( \mathbf{w}^{(j)}\Big\vert \omega^{(j)} \right).
\end{align*}
Let $\mathbf{f}_{t}^{(j)}$ and $\mathbf{z}_t^{(j)}$ denote the forward probability, and the state indicator vector on  day $t$ for household $j$, respectively. The probability of observing the data in each household is
\begin{align}
P\left(\mathbf{w}^{(j)}|\omega^{(j)}\right)&=P\left(w_{1}\right)\prod_{t=2}^{\omega^{(j)}} P\left(w_t\vert\mathbf{w}_{(1:t-1)}\right)\nonumber\\
&=\left(u\cdot \mathbf{z}_1^{(j)}\right) \prod_{t=2}^{\omega^{(j)}} \left(\mathbf{f}_{t-1}^{(j)}e^Q\right)\cdot \mathbf{z}_t^{(j)}, \nonumber
\end{align}
 that is, the within-household likelihood is a product of the normalising constants for the forward probabilities. Thus the within-household likelihood is calculated as a by-product of the expectation calculations.

\section{Results}
\label{sec:results}
Our inference methods are compared against 16 simulations with true parameter values $(\alpha,\beta,\gamma)=(0.32,0.4,1/3)$ and $50,000$ households of size $N=3$ (the average household size in Australia is estimated to be 2.6 \citep{ABS:2011}). These parameters are chosen such that the average infectious period is three days (this is a typical infectious period for influenza), $R_0=\beta/\gamma=1.2$ and $R_*\approx 1.8$. For the branching process  approximation (BPA) the number of realisations used to estimate the distribution of $G$ was $10^3$. \\

Each algorithm is based upon a Bayesian Markov chain Monte Carlo (MCMC) framework in order to estimate the joint posterior distribution of our parameters \citep{Gilks1995}. In particular, the BPA is a Metropolis-Hastings algorithm and the DA MCMC is a single-component Metropolis-Hastings algorithm. Each algorithm is run at various stages of the epidemic in order to show how the posterior distributions converge as more households become infected; the inference for each simulation is run after 50, 100, 200, 300 and 400 households become infected. For the BPA, for each simulation, at each stage of the epidemic, $10^5$ MCMC samples are obtained with a burn-in of 1000 iterations. For the DA MCMC, for each simulation, at each stage of the epidemic, $2.5\times10^6$ samples are obtained with a burn-in of $10^6$ iterations respectively. Both algorithms are implemented with prior distribution for $\left(\alpha,\frac{\beta}{\gamma},\frac{1}{\gamma}\right)$ of $U(0.05,1)\times U(0.25,4)\times U(0.25,7)$. The BPA was implemented with a
\begin{equation}
X|Y\sim N\left(Y,
\begin{bmatrix}
       0.01 & 0 & 0           \\[0.1em]
       0 & 0.02           & 0 \\[0.1em]
       0           & 0 & 0.05
     \end{bmatrix}\right) \nonumber
\end{equation}
proposal distribution. The DA MCMC is implemented by proposing moves (i)-(v) with probabilities $q_1=q_2=0.05$ and $q_3=q_4=q_5=0.3$ respectively.
Our results are displayed in terms of maximum a posteriori (MAP) estimates of the model parameters, $(\alpha,\beta,\gamma)$, and MAP estimates of key epidemiological parameters $(R_*,r)$ and joint posterior density estimates of $(R_*,r)$. All kernel densities were estimated using the freely available MATLAB packages {\tt kde2.m} and {\tt akde.m} \citep{Botev:2010}. \\

\begin{figure}[tb]
	\centering
	\includegraphics[width=\columnwidth]{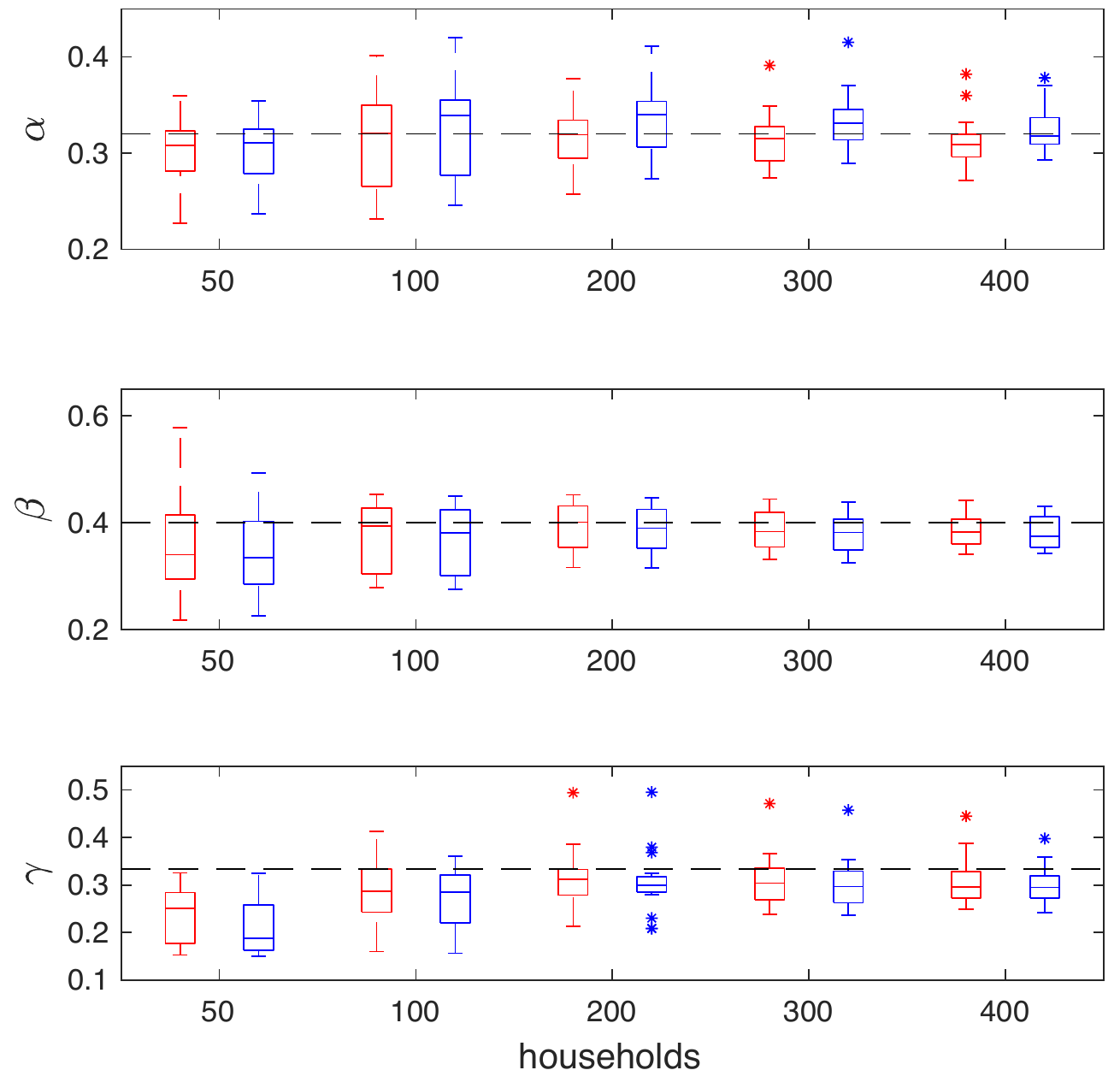}
	\caption{ Boxplots of maximum a posteriori (MAP) estimates of $(\alpha,\beta,\gamma)$ from 16 simulations. Blue and Red boxes correspond to results from $10^5$ iterations of the BPA and $2.5\times10^6$ iterations of the DA MCMC algorithm respectively.  MAP's are calculated from 3 dimensional kernel density estimates. The pairs of boxes from left to right are MAP's from inference based upon data with 50, 100, 200, 300 infected households. Black dotted lines indicate the true parameter values at $(\alpha,\beta,\gamma)=(0.32,0.4,1/3)$.}
	\label{fig:boxes1}
\end{figure}

\begin{figure}[tb]
	\centering
	\includegraphics[width=\columnwidth]{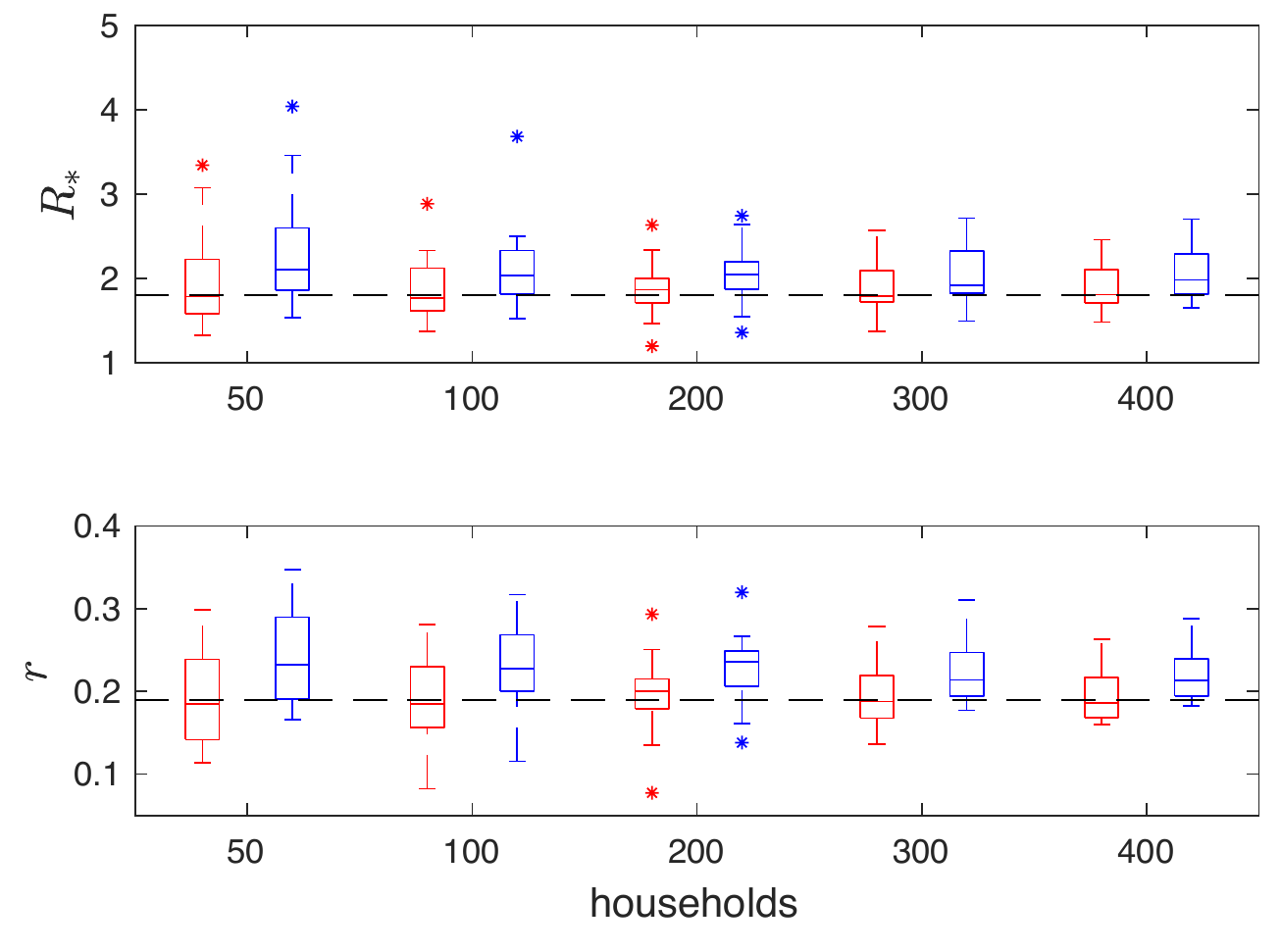}

	\caption{ Boxplots of maximum a posteriori (MAP) estimates of $(R_*,r)$ from 16 simulations. Blue and Red boxes correspond to results from $10^5$ iterations of the BPA and $2.5\times10^6$ iterations of the DA MCMC algorithm respectively. MAP's are calculated from 2 dimensional kernel density estimates. The pairs of boxes from left to right are MAP's from inference based upon data with 50, 100, 200, 300 and 400 infected households. Black dotted lines indicate the true parameter values at $(R_*,r)\approx(1.803,0.190)$.}
	\label{fig:boxes2}
\end{figure}

\begin{figure*}[tb]
	\centering
	\includegraphics[width=2\columnwidth]{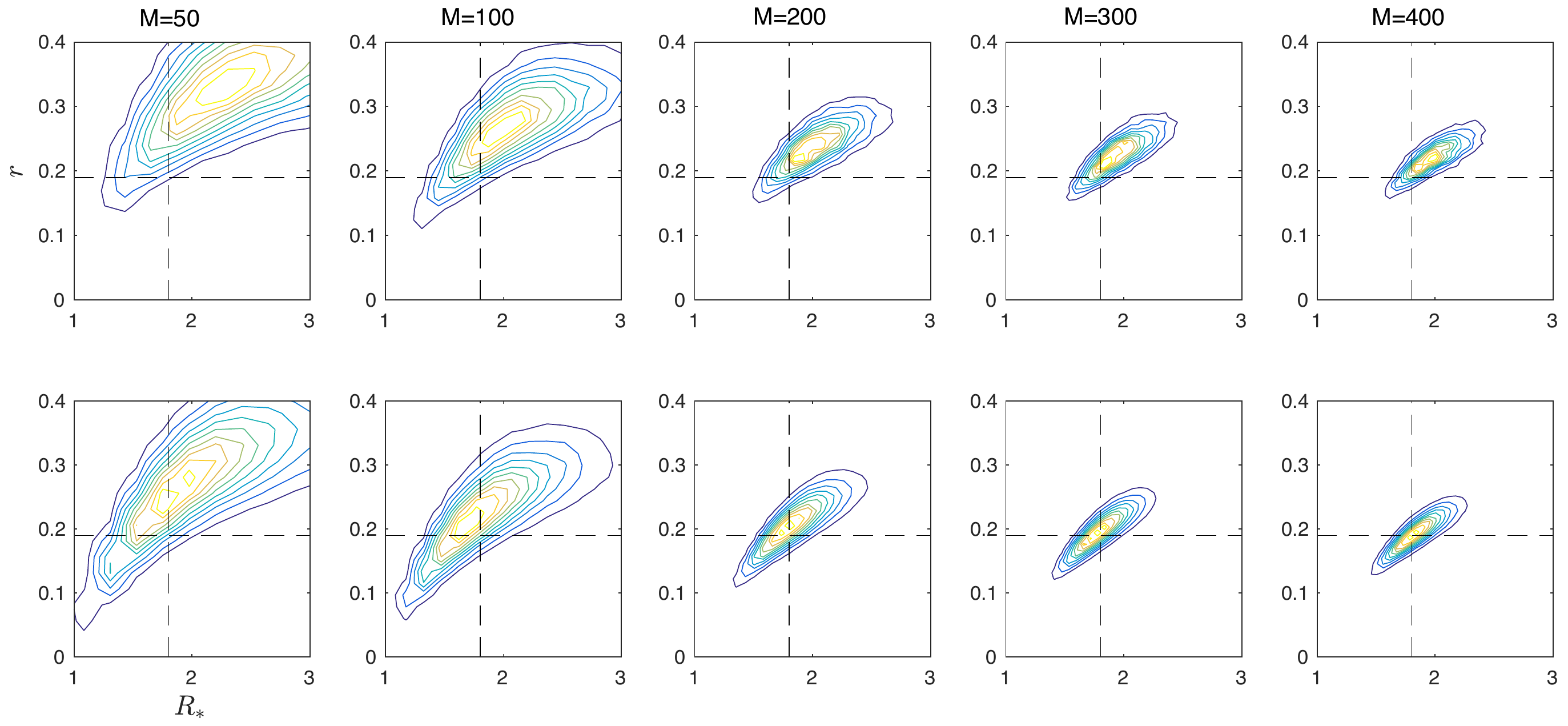}
	\caption{Contour plots of the joint posterior density of $R_*$ and $r$ from a single simulation. The top and bottom panels are results from $10^5$ iterations of the BPA and $2.5\times 10^6$ iterations of the DA MCMC algorithm respectively. The panels from left to right are posteriors from inference based upon data with 50, 100, 200, 300 and 400 infected households. The intersection of the black dotted lines indicate the true parameter values at $(R_*,r)\approx(1.803,0.190)$.}
	\label{fig:jointdensity}
\end{figure*}

Figures \ref{fig:boxes1}-\ref{fig:jointdensity} show a great deal of similarity between the approximate and the exact results, with a mild but noticeable difference in the posterior distributions in Figure \ref{fig:jointdensity}, particularly when only 50 or 100 households are infected. In particular, it appears that the BPA results overestimate $r$ compared to the DA MCMC results in the earliest stages of the epidemic. From Figure \ref{fig:boxes1}, we observe that MAP estimates begin negatively biased for all parameters and converge towards the true parameter values as more data is obtained. For all parameters the variability of the MAP estimates appear to decrease as more data is obtained. It appears that the MAP estimates of $\alpha$ and $\beta$ from the BPA method are slightly positively and negatively biased respectively. It should be noted that these box plots make the correlation structure of the parameters unclear; this is not presented here as the dimension of the parameter space makes the correlation structure difficult to display. In Figure \ref{fig:boxes2} we observe that both $r$ and $R_*$ estimated with little bias by the DA MCMC method but are slightly positively biased by the BPA method. The variance in MAP estimates in Figures \ref{fig:boxes1} decrease as more households become infected. The posterior density estimates of $(R_*,r)$ for a single simulation, displayed in Figure \ref{fig:jointdensity}, shows that the posterior density initially has large variability, but the variability decreases as more households become infected.\\

\begin{figure}[tb]
	\centering
	\includegraphics[width=\columnwidth]{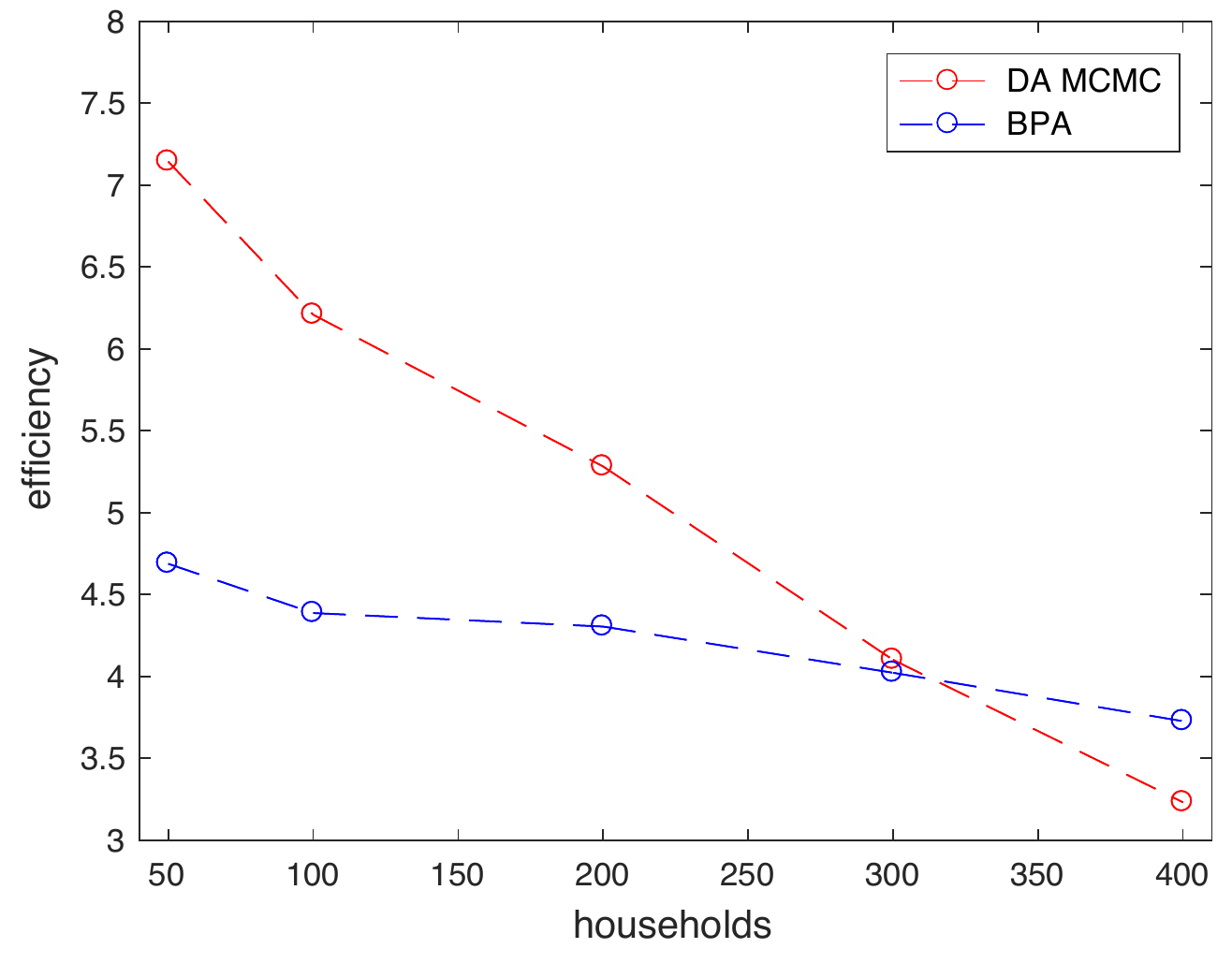}
	\caption{Plots of the efficiency of each method against the number of infected households. Here efficiency is presented in terms of  log multivariate effective sample size per minute. These estimates are based upon running each algorithm for a single simulation with  50, 100, 200, 300 and 400 infected households.}
	\label{fig:efficiency}
\end{figure}

The efficiency of the two algorithms cannot be compared directly in terms of iterations per time, as samples from the DA MCMC are more highly correlated than samples from the BPA \citep{Pooley_2015}. Hence, the algorithms are compared in terms of their multivariate effective sample size per minute, where the multivariate effective sample size is an estimate of the number of independent samples in a dataset \cite{Vats:2016}.
Figure \ref{fig:efficiency} shows that the DA MCMC is initially much more efficient than the BPA algorithm, however it scales poorly as more data is obtained and is less efficient than the BPA after 400 households are infected.

\section{Discussion}
\label{sec:discussion}

In this paper we have implemented a DA MCMC algorithm for exact inference on a stochastic SIR household model and derived a method to approximate the likelihood for an SIR household branching process. These allow us to perform Bayesian MCMC inference to compute posteriors for both $R_*$ and $r$, which are of importance for public health planning. The posteriors from both methods are very similar, showing that the BPA is accurate. These posteriors show good convergence to the true parameter values---this indicates that the FF100 study data can be highly informative despite the amount of missing information. In particular, posterior densities exhibit little variability after 200 households have become infectious. We compared the efficiency of the methods and found that the DA MCMC is superior for data with up to 300 infected households, however due to the poor scaling properties of the DA MCMC, the BPA is preferred for data of more than 300 infected households. However the BPA may introduce some slight positive bias to estimates of $R_*$ and $r$. It should be noted that due to computation time we have only presented results from 16 simulations and a single parameter set. This is the first study that we are aware of to estimate these quantities using early epidemic data and a Bayesian framework.\\

Another way to analyse the stochastic SIR household model is to cast it as a multi-type branching process \citep{Athreya:1972,Dorman:2004}. The theory of these is well developed and hence we can write down equations for many of the quantities we need in order to calculate a likelihood, but actually solving these is too inefficient for practical inference where the likelihood calculation is embedded in an MCMC scheme and hence needs to be repeated many times. Indeed the equations for the probability generating function for the full branching process are very simply stated, but their solution involves a multi-dimensional inversion \citep{AW92}. As such, the approach we have taken with the BPA is to factor the likelihood in a non-standard way, using a small number of well motivated assumptions. Our factorisation allows us to calculate its parts using a combination of numerical techniques; in particular, matrix exponential methods (Section \ref{sec:within}) as well as stochastic simulations and numerical convolutions (Section \ref{sec:sgh}). Each method is appropriate for the task and relatively efficient. For example the simulation to calculate the distribution of $G$ can be programmed efficiently as there is no conditioning involved, so all the generated realisations can be used.\\

Still, each method is relatively expensive to run for larger data sets, and there is room for improvement in many aspects of the procedure. In particular, no attempt has been made to parallelise any part of the BPA algorithm. This would be relatively trivial as most of the calculations are independent of each other and hence this would provide a large speed-up. For example, the simulations to calculate $G$ and the expectation calculations within each household could be parallelised. In \cite{Black:2016}, we also used a tree data structure to minimise the number of operations needed to calculate the within-household likelihood. A similar approach could be taken here to minimise the cost of the expectation calculations, as well as casting them as explicit path integrals that can typically be solved more efficiently \citep{PS02}. Another aspect of our algorithm that can be tuned is the time step used in the numerical integrations. Decreasing this will result in a faster running time, but a larger error in our final posteriors.
While the DA MCMC algorithm may not be parallelisable, efforts could be made to optimise the move proposal density, $\{q_1,\dots,q_5\}$, or to use proposals informed by the model \citep{Pooley_2015,Fintzi:2017}. \\

There are a number of extensions that could be made quite easily to this methodology. An exposed period can be added to the model, but this changes the processes somewhat in that households are no longer observed at the time of first infection, but after the first individual becomes infectious (and displays symptoms). Thus we would need to track the distribution of exposed but not yet infectious households. One aspect that becomes easier, for the BPA method, with the addition of an exposed period is that longer chains of household infections become less likely on a given day. If the exposed period is sufficiently long with high enough probability (say, typically greater than 1 day) then we can approximate the distribution of newly exposed households with just a single generation. The efficiency of the DA MCMC method is likely to scale much worse for an SEIR model, as there will be much more missing information to sample. Another extension would be the incorporation of a realistic distribution of household sizes within the population. For the BPA, the expectation calculations would essentially remain the same, but the proportions of each size of household would need to be taken into account in the between-household likelihood calculation, in the simulations and the convolution procedure. For the DA MCMC these proportions will need to be accounted for in the augmented likelihood and acceptance probabilities.\\

The biggest weakness of this work is that we assume perfect detection of infectious cases. Especially for diseases such as influenza, there can be a large fraction of asymptomatic cases and hence partial detection is the best that can be achieved. In previous work we have not made this perfect detection assumption, but instead assumed that there is some probability per case of detection \citep{Black:2016}. Many aspects of this work could be extended to incorporate partial detection, but the largest challenge for the BPA is modelling the distribution of currently unobserved households and how they contribute to the overall force of infection. The convolution approach may be appropriate here, especially given how fast this is using modern GPU hardware, but this is a topic for further research. This also becomes challenging for the DA MCMC approach, as the missing data can be very high dimensional as it samples from parameter space corresponding to low observation probabilities. \\

\begin{acknowledgements}
JNW acknowledges the support provided by the Dr Michael and Heather Dunne PhD Scholarship in Applied Mathematics.
JVR acknowledges the support of the ARC (Future Fellowship FT130100254). AJB was supported by an ARC DECRA (DE160100690). All authors also acknowledge support from both the ARC Centre of Excellence for Mathematical and Statistical Frontiers (CoE ACEMS), and the Australian Government NHMRC Centre for Research Excellence in Policy Relevant Infectious diseases Simulation and Mathematical Modelling (CRE PRISM$^2$).
\end{acknowledgements}

\bibliographystyle{spbasic}      
\bibliography{adelaide_refs.bib}   


\end{document}